\documentclass[manuscript,screen]{acmart}
\usepackage[utf8]{inputenc}
\usepackage{xcolor}
\usepackage{url}
\usepackage{amsmath}

\usepackage{graphicx}

\usepackage{tabularx}
\usepackage{multirow}
\newcommand*\rot{\rotatebox[origin=c]{90}}

\AtBeginDocument{%
  \providecommand\BibTeX{{%
    \normalfont B\kern-0.5em{\scshape i\kern-0.25em b}\kern-0.8em\TeX}}}

\title{Evaluating Bayesian Model Visualisations}

\author{Sebastian Stein}
\orcid{0000-0003-1828-4008}
\email{sebastian.stein@glasgow.ac.uk}

\author{John H. Williamson}
\email{johnh.williamson@glasgow.ac.uk}

\affiliation{%
    \department{School of Computing Science}
    \institution{University of Glasgow} \city{Glasgow}
    \state{Scotland}
    \country{United Kingdom}
}


\begin{abstract}
    Probabilistic models inform an increasingly broad range of business and policy decisions ultimately made by people. Recent algorithmic, computational, and software framework development progress facilitate the proliferation of Bayesian probabilistic models, which characterise unobserved parameters by their joint distribution instead of point estimates. While they can empower decision makers to explore complex queries and to perform what-if-style conditioning in theory, suitable visualisations and interactive tools are needed to maximise users' comprehension and rational decision making under uncertainty. In this paper, propose a protocol for quantitative evaluation of Bayesian model visualisations and introduce a software framework implementing this protocol to support standardisation in evaluation practice and facilitate reproducibility. We illustrate the evaluation and analysis workflow on a user study that explores whether making Boxplots and Hypothetical Outcome Plots interactive can increase comprehension or rationality and conclude with design guidelines for researchers looking to conduct similar studies in the future.
\end{abstract}

\begin{teaserfigure}
    \centering
    \includegraphics[width=\textwidth]{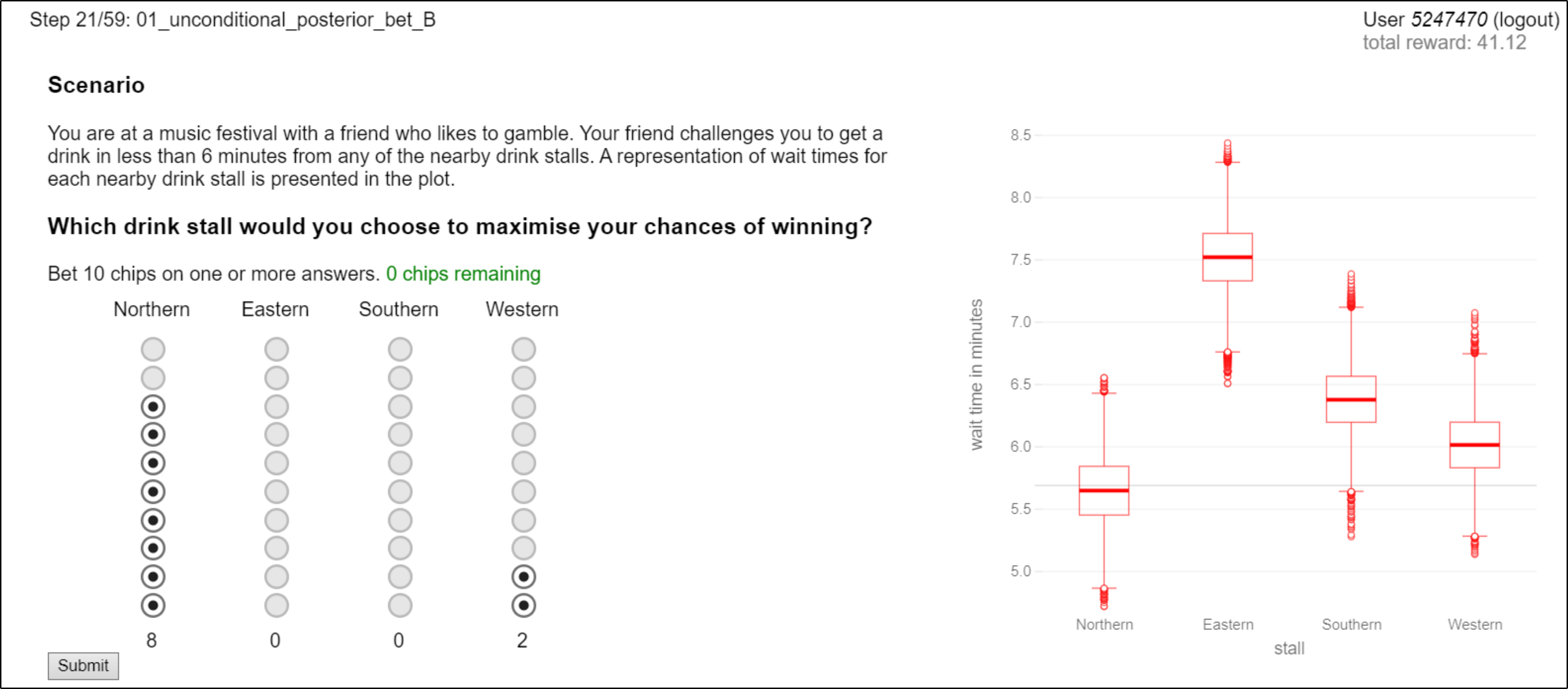}
    \caption{Evaluation UI supporting standardised task-based evaluation of Bayesian Model Visualisations. It includes the task  \emph{Context}, the \emph{Query} to be answered, the \emph{Answer Input}, and \emph{Acknowledgement} on the left, and the \emph{Model Visualisation} on the right. The specific answer input widget and model visualisation are generated dynamically from the task specification. The \emph{MultiBet} answer input is proposed here for capturing distributions of decisions. The Interactive Boxplot updates based on - possibly conditional - resampling from the Bayesian model back-end, triggered by user interactions.}
    \label{fig:task_ui}
    \Description{A UI screenshot during an example task displayed using the proposed study framework.} 
\end{teaserfigure}

\begin{document}

\maketitle

\section{Introduction}

We consciously make decisions under uncertainty in everyday situations: dressing appropriately for the weather, leaving sufficiently early to catch the bus, planning for retirement, or interpreting medical test results\cite{kim2017}. Most of us have never been formally taught to process statistical or probabilistic data \cite{sedlmeier2001} and have difficulty reasoning about such data \cite{koller2009,tversky1974, diaz2007}. Inferring conditional probabilities in particular is a challenging cognitive process that requires a deep statistical understanding \cite{lambert2018_ch12} and can have dire consequences if done incorrectly \cite{Gigerenzer2003}.

Data-driven tools that make predictions about the future in support of human decisions are inherently uncertain. Yet, although they may have an internal representation of uncertainty, it is often hidden from the user \cite{fernandes2018}. On the one hand, communicating uncertainty is challenging \cite{Spiegelhalter1393} and can be more confusing than helpful \cite{greis2017}. On the other hand, it has been shown that uncertainty displays can increase trust \cite{kay2013} in a system, improve understanding of the data-generating process \cite{2015-hops, joslyn2013, kay2016}, help align users' mental model with the computational model \cite{kim2017}, explicitly reduce human biases \cite{tsai2011}, and can improve decision-making \cite{fernandes2018, kay2016}. 

Beyond day-to-day decisions, a large portion of scientific research is concerned with solving inverse problems of estimating unobservable parameters from noisy data and quantifying the uncertainty in those estimates. Some common data science practices have been criticised \cite{kay2016b} for being a unidirectional, statistician-controlled process: starting with a dataset that is publicly available or has been collected in advance, models are learned and compared and a subset of evaluation data is presented via static numerical and visual representations. There is limited scope for users to probe: whether a representation accurately captures domain knowledge; sensitivity to specific data points; uncertainty in predictions; or alternative explanations of the same observations. The ability to explore through interaction becomes increasingly important as models form the basis for human decision-making as well as our scientific understanding.

Bayesian probabilistic models have received significant attention as they capture uncertainty in the data, latent variables, and predictive posteriors, and domain-knowledge can be incorporated through priors \cite{bayes1763}. Significant progress has been made in Monte Carlo-based inference \citep{spiegelhalter2009}, variational methods, and deep learning-based probabilistic models such as Generative Adversarial Networks \citep{goodfellow2014generative} and invertible neural networks \citep{rezende2016variational}. The software stack supporting this work has also matured with public libraries such as Tensorflow Probability, PyMC \cite{salvatier2016}, Pyro, Stan and WebPPL. Due to their ability to provide calibrated estimates of uncertainty in the unobserved model parameters, Bayesian models are used widely in the scientific community to solve inverse problems, including estimating factors influencing wildfires \cite{silva2015} and the effect of non-medical interventions in epidemiology \cite{ferguson13}.

Interactive tools and visualisations for Bayesian models have the potential to let users explore otherwise opaque alternative explanations for observed data, similar to Explorable Multiverse Analysis \cite{dragicevic2019}, and support cognitively difficult processes such as conditioning \cite{tsai2011, taka2020, micallef2012}.

While novel uncertainty visualisations, Bayesian model visualisations, and interactive visualisations are being proposed continually, rigorous evaluations are lacking for some and are narrow in application focus or query type for others \cite{2015-hops, kay2016, cole1989}. Overall, the diversity in evaluation protocols employed across studies makes it difficult to compare results, replicate experiments \cite{micallef2012}, and synthesise generalizable conclusions through meta-reviews \cite{hullman_error}. Studies that do provide a quantitative user study evaluating Bayesian model visualisations and visualisations of uncertainty more generally differ in aspects such as the examined behavioral target, the subset of (Bayesian model) query types, the way user responses are elicited, and the chosen cost function applied to sub-optimal responses, among other factors \cite{hullman_error}. As a consequence, the fields progress is difficult to gauge, and some previously reported effects are irreproducible \cite{micallef2012}. A robust, standardised, evaluation framework to quantify the effect of visualisation on Bayesian model interpretation (comprehension, predictability) and on decision making (rationality) would provide several benefits: data and evaluation results could be aggregated across individual studies, building larger more diverse samples than any individual study could feasibly collect; results obtained with different visualisation types could be more easily benchmarked, akin to benchmarking new machine learning algorithms on public datasets; barriers for new researchers to enter the field would be reduced by providing clear, specific guidelines on how to evaluate new combinations of models and visualisations.

As a contribution to the ongoing methodological renaissance \cite{nelson2018} in human-subject research addressing the recent replication crisis, this paper aims to further the community discussion of evaluation methodology and
\begin{itemize}
    \item proposes a common evaluation protocol for evaluating Bayesian model visualisations (Section \ref{sec:evaluation_protocol})
    \item introduces a publicly available software framework that implements this protocol (Section \ref{sec:software_framework}), and
    \item illustrates the analysis workflow on a user study comparing static, animated, non-interactive and interactive uncertainty visualisations (Sections \ref{sec:case_study}-\ref{sec:analysis}).
\end{itemize}

\section{Related Work}

\subsection{Bayesian probabilistic models}
\label{sec:bayesian_models}

Let's briefly introduce the notation for Bayesian statistics used throughout this paper. A Bayesian probabilistic model factorises the joint distribution $p(x, \theta)$ of observable random variables $x \in \mathbb{R}^D$ and unobserved (latent) random variables $\theta \in \mathbb{R}^{D'}$ into a \emph{prior} distribution $p(\theta)$ representing epistemological uncertainty and a \emph{likelihood} $p(x | \theta)$ representing aleatory uncertainty \eqref{eq:joint}. Usually, $x$ corresponds to data one can collect and $\theta$ represents the parameters of a statistical model. Bayes' Theorem expresses how one should rationally update ones \emph{posterior} belief about latent variables in light of observations $X$ \eqref{eq:Bayes}. The \emph{posterior predictive} distribution defines what one should rationally believe about future events $x$ given past experience \eqref{eq:posterior_predictive}. The integrals in \eqref{eq:Bayes} and \eqref{eq:posterior_predictive} are often intractable and are therefore either approximated using Markov Chain Monte Carlo or variational inference, or avoided in likelihood-free methods like GANs. These methods have in common that distributions $p(x, \theta)$ and $p(x, \theta|X)$ can be represented by sets of samples $(x^i, \theta^i)$. The prior, prior predictive, posterior, and posterior predictive distributions can be high dimensional with complex covariance structure and we currently lack tools for decision-makers and domain experts with potentially little statistical background to interact with them effectively.

\begin{align}
    p(x, \theta) &= p(x| \theta) p(\theta) \label{eq:joint}\\
    p(\theta| X) &= \frac{p(X| \theta) p(\theta)}{\int p(X|\theta)p(\theta) d\theta} \label{eq:Bayes}\\
    p(x|X) &= \int p(x|\theta)p(\theta|X) d\theta \label{eq:posterior_predictive}
\end{align}

\subsection{Bayesian model visualisations}

A widely studied problem in the literature on Bayesian decision making is commonly known as the mammography problem \cite{eddy_1982}, in which participants are asked to estimate the probability of a hypothetical patient having breast cancer, given a positive mammography result and estimates for the base rate and the test's specificity and sensitivity. In this example, the test result is an observed variable and the presence of cancer is a latent variable. Solving this problem requires participants to apply Bayes theorem \eqref{eq:Bayes}. \citet{eddy_1982} found that 95 out of 100 physicians estimate the wrong quantity and, as a result, greatly overestimate the probability of the presence of cancer. \citet{cole1989} observed improved performance where participants used visual aids, i.e. contingency tables, signal detection curves, detection bars or probability maps when solving this problem. A large-scale online crowd-sourced replication study \citep{micallef2012} comparing area-proportional Euler diagrams, glyph representations, and hybrid diagrams combining both was unable to confirm previously observed benefits of presenting a visual representation alongside text, only observing increased performance over the text-only condition when numerical information was removed from the associated text. While it is still an open research question how best to present visual information in this particular problem, it is only representative of a small subset of queries one can answer with probabilistic models. How accurately would participants solve a related problem in which base rates and test errors were uncertain, test outcomes were continuous or multi-dimensional, or the question was rephrased to ask which latent state (cancer or no cancer) is most likely given the test result, or how confident they are that \emph{no cancer} was most likely? In this paper, we map the space of possible queries one can answer with probabilistic models and argue that more research is needed to explore how visualisations can support users across the whole query space.

Inspiration can be drawn from generic visualisations of uncertainty such as boxplots, density plots, Quantile Dot Plots \cite{kay2016}, Hypothetical Outcome Plots (HOPs) \cite{2015-hops}. It is unclear how well these approaches scale up to the large number of potentially high-dimensional variables commonly found in Bayesian models. One particular challenge is how to communicate complex correlation structure across variables when only marginal distributions are displayed. \citet{taka2020} review available visualisation tools for Bayesian analysis. Animation may help by sequentially showing, for example individual draws from the joint distribution \cite{2015-hops}. Interactive primitives \cite{taka2020} could also help with exploratory data analysis. In this paper we test the hypotheses that animation or interactivity could help with Bayesian decision-making with a user study to illustrate the analysis workflow for the proposed evaluation protocol.

\subsection{Evaluating uncertainty visualisations}

In a recent review of evaluation protocols in uncertainty visualisation, \citet{hullman_error} highlighted the diversity in design decisions made by different authors, encouraged a less ad-hoc approach to study design, and more explicit reporting of design decisions. Specifically, they distinguish design decisions at six different levels, including the behavioral target (e.g. performance or quality of user experience), the expected effect (e.g., accuracy or satisfaction), the evaluation goal (e.g. comparison of different visualisations or understanding why a visualisation works well), measures of users' behavior (e.g., a decision or self-reported satisfaction), elicitation (how is user behavior captured?), and analysis. With different decisions made at each of these levels affecting the conclusions drawn from individual studies, we argue that there is a need for the community to converge at a small subset of possible evaluation pathways to foster reproducibility, support benchmarking, and enable meta-studies. \citet{breslav2014} proposed a software system that records users' micro-interactions when interacting with uncertainty visualisations and to help designers analyse user behavior. In this paper, we propose one evaluation pathway for evaluating Bayesian model visualisations with respect to users' rationality in decision-making and users' comprehension of the uncertain information represented by the model. We specifically follow recommendations in \cite{hullman_error} to use decision frameworks for realism and control by measuring the quality of decisions using rational choice theory, to calibrate the user by providing feedback with regards to the decision quality and accuracy of non-decision responses, and by designing the MultiBet input element that allows participants to spreading bets across multiple outcomes where committing to a single decision is difficult. We aim to encourage reuse of this specific pathway by contributing a software framework that can be readily reused by the research community. 

\section{Evaluation Protocol}
\label{sec:evaluation_protocol}

In presenting the evaluation protocol and subsequent user study we follow the taxonomy proposed in \cite{hullman_error}. In this Section, we discuss behavioral targets, measures of user behavior and elicitation techniques as they are proposed to be fixed for evaluating user rationality and comprehension of Bayesian models and are implemented in the proposed software framework. The expected effects, evaluation goals, and analysis of a specific user study using this protocol are detailed in Section \ref{sec:case_study}.

\subsection{Query Space for Bayesian Models}
\label{sec:query_space}

Prior to specifying behavioral targets, it is useful to explore the space of possible questions users could answer with the help of a Bayesian model. In Section \ref{sec:bayesian_models} we described Bayesian models as a factorized representations of $p(x, \theta)$ and $p(x, \theta|X)$, inviting questions about both \emph{prior} and \emph{posterior} distributions in the presence of observed data. In both cases, Bayes' theorem also provides a solution for conditioning on a subset of variables. Their factorization suggests another dimension in the space of possible questions specifying whether a question relates to \emph{observable} or \emph{latent} variables. For each variable the model holds information about multiple quantities: the \emph{identitity} (or index) of a specific variable, what \emph{values} could a variable take on, and how \emph{confident} is the model that a variable might take on a particular value.

Evaluations of Bayesian model visualisations can therefore explore a high-dimensional space of user questions, formed by the outer product of what the model can tell us
\begin{itemize}
    \item in the presence or absence of data?
    \item when incorporating additional information about a subset of variables?
    \item about observable and latent variables?
    \item about a variables identity, value, or it's confidence in a variable's value?
\end{itemize}

\begin{table}[]
    \centering
    \begin{tabular}{l|l||l|l}
         & & \multirow{5}{*}{\rot{Observability}} & \multirow{5}{*}{\rot{Quantity}}\\ 
         & & \\
         & & \\ 
         & & \\
         Query & Numerical Objective & & \\
         \hline
         
         \begin{minipage}[t]{0.5\columnwidth}
         What return do you expect at least from asset $i$ with $95\%$ confidence?
         \end{minipage}
         & $ \operatorname{argmin}_\tau \, |P(x_i \geq \tau) - 0.95| $ & $x$ & value \\[12pt]
         
         \begin{minipage}[t]{0.5\columnwidth}
         Which observed return would make it most likely that asset $i$ generated it? 
         \end{minipage} &
         $\operatorname{argmax}_{\tau}\, p(\theta_i|x_i=\tau)$ & $\theta$  & value\\
         
         \begin{minipage}[t]{0.5\columnwidth}
         Which asset would you buy to maximize expected investment return? 
         \end{minipage}
         & $\operatorname{argmax}_i \, \mathbb{E}_{p(x, \theta)}[u(x_i)]$& $x$ & id \\[12pt] 
         
         Which asset is most likely to return at least $2\%$ & $\operatorname{argmax}_i \, P(x_i \geq 0.02) $ & $\theta$ & id. \\[2pt]
         
         \begin{minipage}[t]{0.5\columnwidth}
         How confident are you that asset $i$ will return at least $2\%$?
         \end{minipage}
         & $ P(x_i\geq 0.02) $& $x$ & conf. \\[2pt] 
         
         \begin{minipage}[t]{0.5\columnwidth}
         How confident are you that an observed return of $2\%$ was most likely generated by asset $i$?
         \end{minipage} 
         & $p(\theta_i| x_i=0.02)$ & $\theta$  & conf. \\[12pt] 
    \end{tabular}
    \caption{Example queries at different points along the Observability and Quantity dimensions of the probabilistic model query space. Each query corresponds to a numerical objective, which enables quantitative assessment of users' responses.}
    \label{tab:example_queries}
    \Description{Example queries at different points along the Observability and Quantity dimensions of the probabilistic model query space. Each query corresponds to a numerical objective, which enables quantitative assessment of users' responses.} 
\end{table}

To illustrate the diversity of possible model queries, we consider the problem of investing a fixed sum of money for a fixed duration into one of a discrete set of assets. Asset returns could be represented with a Bayesian model that specifies the realised return as observable $x$ and asset classes as latent variables $\theta_i$. Example queries representing different points in the query space along the Observability and Quantity dimensions are shown in Table \ref{tab:example_queries}. 

We propose this mapping as a tool to support systematic design of user tasks for quantitative user studies and as a communication tool to support research reproducibility. A user study aiming to quantify the effectiveness of a universal visualisation tool for Bayesian models should include user tasks that uniformly cover the query space along each of the described dimensions. Different user groups such as domain experts, statistical model builders, and decision makers may only be interested in a smaller query subspace corresponding to their contribution to the Bayesian workflow \cite{betancourt2018}, and evaluation of specialized visualisation tools may focus on user tasks that cover the most relevant subspace for their user group. This subspace could be determined, for example, through an exploratory study with participants from this group that systematically probes the relevance of queries along each of the query space dimensions. Similarly, a visualisation tool that supports a specialized application such as prior elicitation should be evaluated specifically on user tasks querying the probability and value of observable and latent variables prior to observing any data.

\subsection{Behavioral Targets and Measures}

In evaluating Bayesian model visualisations we seek to explore their effect on users' performance, specifically on users' ability to make optimal decisions and on users' understanding of the model. We apply rational choice theory to evaluate decision making and thus refer to the optimality of users' decisions as \emph{rationality} and their level of understanding as model \emph{comprehension}, respectively.

Rational choice theory assumes that user preferences can be expressed as scalar utility functions $u(x)$ over all options of a choice set. If $u(x)$ is known the value of a decision can be quantified as the expected utility under the model \eqref{eq:expected_utility}.

\begin{equation}
    v(x) = \mathbb{E}_{p(x|X)}[u(x)] = \int u(x)p(x|\theta)p(\theta|X) d\theta \label{eq:expected_utility}
\end{equation}

In practice, the utility function is often intrinsic, possibly unique to each user, and not immediately accessible for quantitative evaluation. An extrinsic utility function could be provided as part of the user task in the form of hypothetical betting odds, for example. Participants' incentives could be aligned with an extrinsic utility by asking participants explicitly to maximise utility or by tying reward mechanisms and a portion of participant compensation to expected utility.

Model \emph{comprehension} can be quantified by comparing users' estimates of a modeled quantity to model estimates. Each numerical estimate provided by the user - of confidence, of a variable's value, or of a variable's identity - entails a probability distribution $\hat{p}$. Its divergence from the probability distribution $p^*$ entailed by the optimal response can be measured using the Kulback-Leibler (KL) divergence \eqref{eq:KL}, or a symmetric variant: $D(p^*\, \| \, \hat{p}) + D(\hat{p} \, \| \, p^*)$. The KL-divergence is zero for an optimal response and increases with relative entropy. It has been used previously to evaluate model comprehension in the context of uncertainty visualisations, where users were asked to graphically predict the distributions of events \cite{hullman2018}. Micallef et al.  \cite{micallef2012} proposed the absolute bias $|\hat{p} - p^*|$ as an objective function, which penalizes larger errors more gracefully, whereas Tsai et al. \cite{tsai2011} used a hard binary criterion.

\begin{align}
    D\left (p^*\, \middle\|\, \hat{p} \right) = \int_{-\infty}^{\infty}p^* \, \log\left ( \frac{p^*(x)}{\hat{p}}\right ) dx
    \label{eq:KL}
\end{align}

Equipped with objective functions that measure \emph{rationality} as the similarity between user decisions and optimal decisions \eqref{eq:expected_utility}, and that measure \emph{comprehension} as the divergence between distributions entailed by user responses and optimal responses \eqref{eq:KL}, we can quantitatively evaluate the effect of model visualisations. By sampling user responses to tasks that cover the relevant query subspace across a range of models under one or more conditions we can quantify \emph{how close} user responses are to the optimal response, and \emph{how significant} differences of responses are, comparing user groups, visualisation types, and user behavior models.


\subsection{Elicitation}

Eliciting quantitative user responses to model queries requires associating an appropriate input widget with each query Quantity. Users are familiar with sliders to input continuous and ordinal values, making this type of input widget appropriate for queries related to user confidence or degree of belief, and potentially queries related to the value of a variable. The identity of an observed or a latent variable, such as the investment asset in Table \ref{tab:example_queries}, is neither continuous nor ordinal in the general case. Commonly used discrete choice input widgets include radio buttons, list selectors, and drop-down selectors. The visual response of radio buttons to user input bears an abstract similarity to placing a betting chip on specific outcomes, akin to playing the casino game of Roulette. Reinforcing this cognitive association through the choice of input widget could help communicate the effect of a user's response, particularly in the context of tasks evaluating rationality.

User tasks for evaluating model comprehension can be readily formulated in such a way that the value of a variable and one's confidence in that value are probed independently. Consider, for example, the questions \emph{What return do you expect at least from asset $i$ with $95\%$ confidence?} and \emph{How confident are you that asset $i$ will return at least $2\%$}. For rationality tasks, this is not the case as the question of confidence relates to a decision, represented by one's input choice instead of a property of the model. This could be resolved, for example, by simultaneously asking users to input their decision through a radio button list and to indicate their confidence in their decision with a slider. A joint response of a decision and confidence level, however, cannot be evaluated quantitatively using expected utility as it only specifies the probability (density) of the action distribution at a single point, leaving the distribution under-specified. Uncertain inputs for ordinal values \cite{greis2017b} in comprehension tasks pose a similar issue for evaluation, despite potentially being more intuitive to use. We therefore propose a new input widget that allows users to \emph{spread} their bets in discrete increments across the choice set.

\begin{figure}
    \centering
    \includegraphics[width=0.33\textwidth]{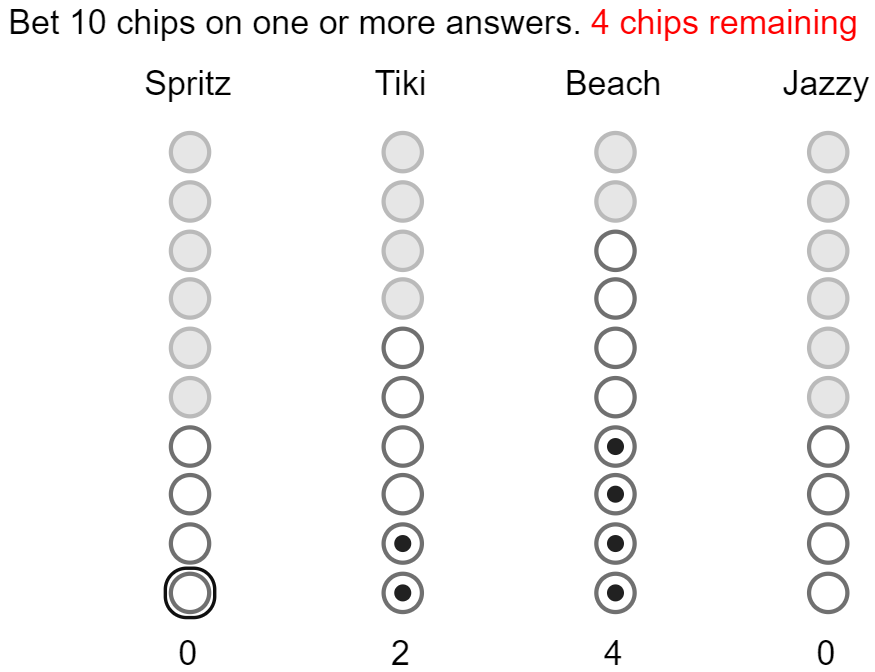}
    \caption{A MultiBet input with a maximum of 10 chips that can be bet on four options. Filled circles represent the current bet, empty circles represent remaining unplaced chips, and gray circles indicate invalid bets given the number of chips that are already placed.}
    \label{fig:multi_bet}
    \Description{A MultiBet input with a maximum of 10 chips that can be bet on four options. Filled circles represent the current bet, empty circles represent remaining unplaced chips, and gray circles indicate invalid bets given the number of chips that are already placed.} 
\end{figure}

The $MultiBet$ input generalizes the radio button list to $M \geq 1$ incremental bets on a set of $N$ options (see Figure \ref{fig:multi_bet}). It consists of $N$ columns with $M$ radio buttons each. Its state $s \in \mathbb{N}_0^N$ with $\left(\sum_i s_i\right) \leq M$ represents the number of metaphorical chips placed on each option. In each column, the bottom-most $s_i$ radio buttons are displayed as selected, akin to a stack of chips placed on top of the bottom edge of the widget. The next $M - \left(\sum_i s_i\right)$ radio buttons above in each column are displayed as deselected, representing the remaining unused budget of chips. All other radio buttons are displayed as disabled, indicating that a bet of more than $s_i + M - \left(\sum_i s_i\right)$ chips on option $i$ is invalid under state $s$. A click on a deselected radio-button in column $i$ increases $s_i$ to the corresponding number of chips and disables additional radio buttons in all columns accordingly. Conversely, a click on a selected radio-button in column $i$ decreases $s_i$ to the corresponding number of chips and enables additional radio buttons in all columns. Note that these interactions update $s_i$ by variable amounts, depending on the specific radio button clicked.


\subsection{Individual Tasks and Sequencing}
\label{sec:individual_tasks}

Each user study task consists of a standardised set of interface elements. The \emph{Query} specifies the question a user is asked to answer by interpreting and potentially interacting with a \emph{Model Visualisation}. Any prerequisites for understanding the query and interpreting the model's visual representation are specified in a task \emph{Context}. An \emph{Answer Input} element, such as a slider or a MultiBet widget, elicits a quantitative user response. An \emph{Acknowledgement} input element such as a standard button widget affords finalizing the user response. Optionally, a \emph{Feedback} element communicates aspects associated with the optimality of the submitted response, such as a decision's expected utility, back to the user.

A comprehensive user study following the proposed evaluation protocol comprises all tasks from the outer product of query types, visualisation types, example models, and levels of task difficulty. This can quickly lead to an unacceptably high completion time per participant: participants of the user study in Section \ref{sec:case_study} completed 24 tasks in 50 minutes, excluding briefing and debriefing. To minimise learning effects as a confounder in user responses, tasks ordering should be randomized across participants. Participants may perceive a heightened cognitive load due to frequent context switches, which may impact their responses if they felt under pressure or they were not provided with the opportunity to take breaks when needed.

\section{Software Framework}
\label{sec:software_framework}

The evaluation protocol described in Section \ref{sec:evaluation_protocol} was implemented as a web-service and is freely available online\footnote{Submitted as supplementary material for double-blind review}. The server is implemented using the Python micro web-framework Flask\footnote{\url{https://flask.palletsprojects.com/en/1.1.x/}}. The client is implemented in JavaScript using React\footnote{\url{https://reactjs.org}} for interactivity, Redux\footnote{\url{https://redux.js.org}} for state management, and D3\footnote{\url{https://d3js.org}} for model visualisations and user feedback. This architecture was chosen to enable remote studies potentially run online and to minimise restrictions on the choice of server operating system or client devices. The server provides a REST API with a unique endpoint for each user study. Study management, participant management, participant response logging and task progress tracking is implemented using SQLite3 for persistence. Participants subscribe to a study on the client upon which a random user ID is generated on the server and added to the study, together with a newly generated randomized task order. At a study and user-specific endpoint, data for the first uncompleted task can be requested. The corresponding user response and other client-side logs are submitted at the same endpoint, which are written to the database and enable the data for the next task to be requested.

Randomized task orderings for each user are generated from a JSON-formatted study template, which contains a nested \emph{TaskList} whose elements are \emph{TaskList}s, \emph{MergeLists} or \emph{Task}s. \emph{TaskList}s specify whether the order of list elements in the template should be preserved or randomized for each user. Nested \emph{TaskLists} can express complex ordering requirements including a fixed ordered introduction followed by randomized tasks followed by a fixed ordered debriefing, and randomized orders of fixed-order sequences. \emph{Task}s define the task-specific info for all interface elements identified in Section \ref{sec:individual_tasks} (\emph{Context}, \emph{Query}, \emph{Model Visualisation}, \emph{Answer Input}, and \emph{Feedback}), and meta-data pertaining to the probabilistic model associated with the task. \emph{MergeList} is a convenience type that specifies two lists of partial tasks from which the outer product is generated; this facilitates the specification of multiple repetitions of tasks where, for example, only the visualisation type or the associated probabilistic model differ across repetitions.

The task specification sent by the server is used by the client to populate a template user interface (UI) layout (see Figure \ref{fig:task_ui}). The template specifies a two-column layout with task \emph{Context}, \emph{Query}, \emph{Answer Input}, and \emph{Acknowledgement} displayed on the left, and the \emph{Model Visualisation} displayed on the right. With interface elements implemented as React components, different \emph{Answer Input} types and \emph{Model Visualisation} types are created dynamically to match the current task's specification. 

The prior $p(x, \theta)$ or posterior $p(x, \theta | X)$ joint distribution of observable and latent variables, represented as Monte Carlo samples, is transmitted separately as a binary blob. A \emph{Generator} object exposes operations on this distribution, such as estimating sample statistics or the marginal (cumulative) density. \emph{Model Visualisation}s interface with the \emph{Generator} and dynamically update their view using D3 to represent properties of the probabilistic model. A Boxplot component, for example, would access the generator's estimates of the median, inter-quartile range and outlier samples for all its relevant variables and dynamically adapts its view to this data. One of the advantages of Bayesian models is the ability to answer what-if style queries expressed by conditioning the joint distribution on additional information. For example, expanding on the set of queries in Table \ref{tab:example_queries}, one can ask \emph{Which asset is most likely to return at least $2\%$ if asset $j$ was to lose in value (to return less than $0\%$)?} by conditioning the predictive distribution on $x_j$ and inspecting $P(x_i \geq 0.02| x_j \le 0)$. Depending on the specificity of the condition there may only be few Monte Carlo samples in the support of the conditional distribution, providing a poor approximation of the density. Requesting additional Monte Carlo samples from the server would introduce significant latency rendering interactive exploration of this class of queries impractical. Instead, we implement bootstrapping on the client by re-sampling a fixed number of MCMC samples with replacement from the support. This approach reduces bias in the local density estimate while keeping latency at a minimum and avoiding the need for closed-loop communication between the client and a server-side Bayesian inference engine. 

\begin{figure}
    \centering
    \includegraphics[width=0.45\textwidth]{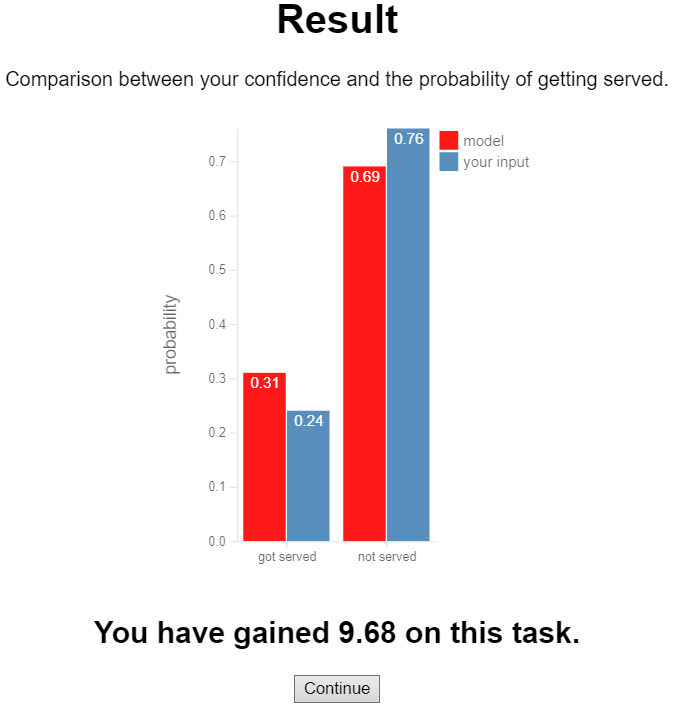}
    \caption{Visual feedback presented to the user immediately after task completion. The probability distribution entailed by the user's response is visually contrasted to the optimal response. The KL-divergence is estimated and re-scaled to a reward in the range $[0, 10]$, where a reward of 10 corresponds to the optimal response. This is done to make it easy and consistent for the user to interpret feedback, and to incentivize users to focus on accuracy.}
    \label{fig:feedback_ui}
    \Description{Visual feedback presented to the user immediately after task completion. The probability distribution entailed by the user's response is visually contrasted to the optimal response. The KL-divergence is estimated and re-scaled to a reward in the range $[0, 10]$, where a reward of 10 corresponds to the optimal response. This is done to make it easy and consistent for the user to interpret feedback, and to incentivize users to focus on accuracy.} 
\end{figure}

The Generator is also used to evaluate user responses with respect to expected utility or its KL-divergence from the optimal response, to provide optional low-latency user feedback. An example of feedback provided to the user immediately after their response was acknowledged is shown Figure \ref{fig:feedback_ui}. Here the probability distribution entailed by the user response is visually contrasted with the same distribution under the optimal response as per the model, and a numerical \emph{Reward} in $[0, 10]$ is given, where 10 corresponds to the optimal response. This is done to make it easy and consistent for the user to interpret feedback, and to incentivize users to optimize their response with respect to the evaluation objective.

All client-side state changes, induced by user interaction or by client-server communication, are implemented as Redux actions. A Redux logger middleware caches all actions locally together with a timestamp. Logs of actions associated with each task are sent together with the user response, the estimate of the objective function, and the reward as a JSON object when the user acknowledges their response, and stored server-side as a string in the SQLite3 database.

Implementing a new user study with this software framework involves providing a study template that specifies task metadata and task order constraints, and optionally implementing new React Components for \emph{Model Visualisations} or \emph{Answer Inputs}. The modular client-side implementation facilitates seamless integration of new UI components, developed by the authors or other researches, in production web-services. In making the software framework that implements the proposed quantitative evaluation protocol publicly available, we hope to reduce the barriers for researchers to evaluate novel visualisations with users, and facilitate research into human perception of uncertainty.

\section{Case Study: Interactive Boxplots and HOPs}\label{sec:case_study}

This Section reports on a case study that aims to quantitatively evaluate two research questions: do animated or interactive visualisations of uncertainty in Bayesian models improve rationality and comprehension over their static and non-interactive equivalents? We apply the proposed evaluation protocol to gather quantitative data that let us answer these questions with a user study for a specific set of static, animated, non-interactive and interactive visualisations.

Static boxplots, error bar charts and confidence intervals have been shown to be regularly misinterpreted \cite{belia2005}, whereas animated Hypothetical Outcome have shown promising results in detecting trends from noisy data Plots \cite{2015-hops}. Interactivity has been shown to help visualise high-dimensional data revealing pertinent low-dimensional views through user-guided projection \cite{faith2007}, align users' cognitive with computational models \cite{kim2017}, and help answer queries requiring Bayesian inference \cite{tsai2011}.

\subsection{Apparatus}

Participants interacted with the client-side web interface described in Section \ref{sec:software_framework} and illustrated in Fig.~\ref{fig:task_ui}. Here, we describe the Bayesian model participants were asked to query and the Model Visualisations chosen to compare the effects of animation and interactivity on user responses.

\subsubsection{A Bayesian model of peak and off-peak queuing delays}

Our objectives for choosing a Bayesian model for this study were that (i) it models a sufficiently complex probability distribution to be representative of a real-world problem (ii) it can be inferred from available real or synthetically generated data (iii) the semantics are easily explained in the briefing session of a user study and (iv) the risk of participants applying previously acquired domain knowledge in answering questions is minimised. We chose to adopt a textbook model from \cite{McElreath2016}, representing the queuing delay across a range of cafes when ordering a coffee, distinguishing between peak and off-peak periods. The data-generating process was modelled hierarchically, assuming that the average peak wait times and differences between average peak and off-peak wait-times in each cafe are drawn from a global joint distribution as in Eqns.~\eqref{eq:model_start}-\eqref{eq:model_end}. A synthetic model with the same structure and latent parameters generated from $ (\sigma_g^{\text{peak}}=1.5, \sigma_g^{\text{diff}}=0.75, \rho=-0.7, \mu_g=(6.5, -1.75)^T, \sigma_x=0.5$) was used to simulate a total of (16 cafes x 2 periods x 5 visits = 160) observations $X$. Hamiltonian Monte Carlo was used to generate 20,000 samples from the joint posterior distribution $p(x, \theta|X)$.

\begin{align}
    \log\, \sigma_g^{\text{peak}} &\sim Normal(0; 0.1) \label{eq:model_start}\\
    \log\, \sigma_g^{\text{diff}} &\sim Normal(0; 0.1)\\
    \tan\, \rho &\sim Normal(0; 1)\\
    \Sigma_g &= \begin{pmatrix}
    (\sigma_g^{peak})^2 & \sigma_g^{\text{peak}}\, \rho\, \sigma_g^{\text{diff}} \\
    \sigma_g^{\text{peak}}\, \rho \,\sigma_g^{\text{diff}} & (\sigma_g^{\text{diff}})^2
    \end{pmatrix}\\
    \mu_g &\sim Normal(\mu=(0,0)^T; \sigma=\mathbb{1})\\
    \mu_i, b_i &\sim Normal(\mu_g, \Sigma_g)\\
    \log\, \sigma_x &= Normal(0; 1)\\
    x_i^{\text{peak}} &\sim Normal(\mu_i; \sigma_x)\\
    x_i^{\lnot \text{peak}} &\sim Normal(\mu_i + b_i; \sigma_x) \label{eq:model_end}
\end{align}

\subsubsection{Model Visualisations}

The joint distribution of queuing delays for a subset of all caffees was represented to participants with one of four different Model Visualisations: Boxplots, Hypothetical Outcome Plots (HOPs), and interactive variants of each. Static boxplots represent the marginal distributions of latent variables and posteriors via aggregate statistics (median, inter-quartile range, outliers etc.). This static view hides covariance, masks distributional anomalies, and it has been shown that boxplots can be difficult to interpret for users with limited training in statistics \cite{belia2005}. In contrast, Hypothetical Outcome Plots have shown to be more readily interpretable particularly with respect to trends across multiple variables \cite{2015-hops}. Hypothetical Outcome Plots (HOPs) represent the joint distribution of latent variables and posteriors via animation by iteratively displaying a single joint sample of all parameters drawn at random. Iterative sampling allows the user to build a mental model of the joint distribution by visually integrating depicted samples over time (time-multiplexing). Sampling gives some access to the model’s covariance. However the user has no means to guide or constrain exploration towards subspaces of interest.

\begin{figure}
    \centering
    \begin{minipage}[t]{0.48\textwidth}
    \includegraphics[width=0.9\textwidth]{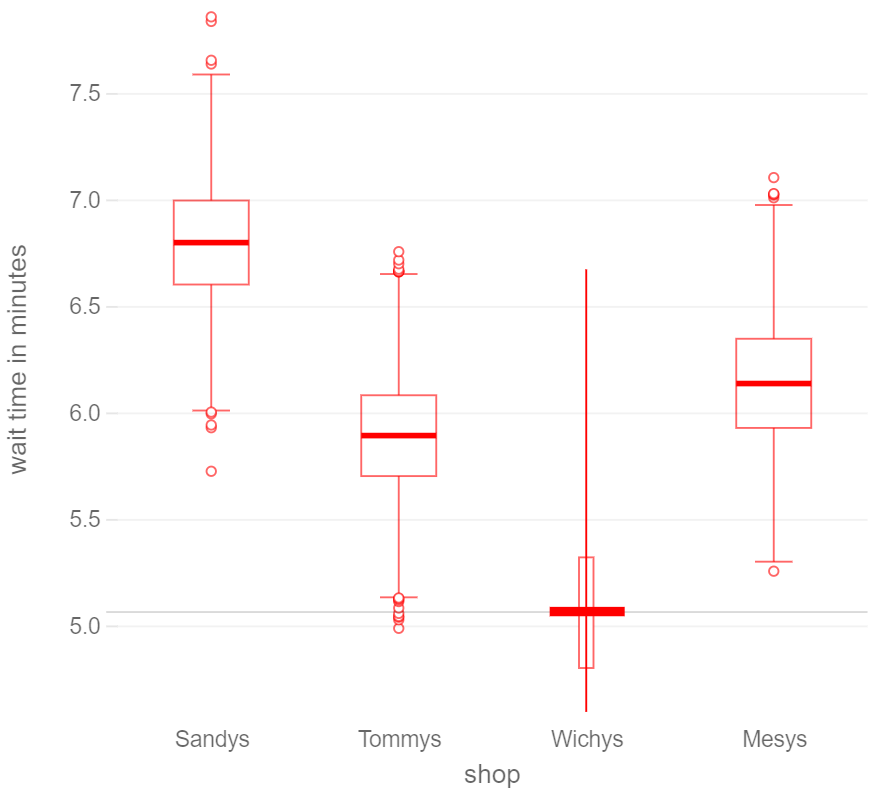}
    \end{minipage}
    \begin{minipage}[t]{0.48\textwidth}
    \hfill
    \includegraphics[width=0.9\textwidth]{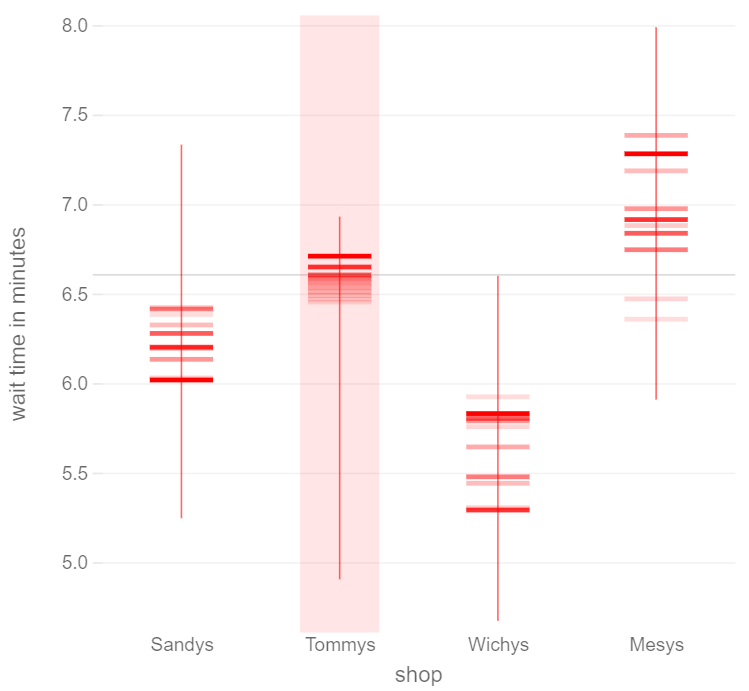}
    \end{minipage}
    
    \caption{Interactive Boxplots (left) enable conditioning of the joint distribution on a small range of a variable of interest, initiated with a mouse click. Here, the user explores the joint distribution under the condition that the wait time at \emph{Wichys} is around 5 minutes. Ballistic Hypothetical Outcome Plots (right) allow the user to explore the joint distribution in a ballistic “pseudo-physical” sequence ordered along a variable of interest. Here, the user clicked on \emph{Tommys} to explore samples from the joint distribution ordered by the wait time at that shop.}
    \label{fig:interactive_vis}
    \Description{Interactive Boxplots (left) enable conditioning of the joint distribution on a small range of a variable of interest, initiated with a mouse click. Here, the user explores the joint distribution under the condition that the wait time at \emph{Wichys} is around 5 minutes. Ballistic Hypothetical Outcome Plots (right) allow the user to explore the joint distribution in a ballistic “pseudo-physical” sequence ordered along a variable of interest. Here, the user clicked on \emph{Tommys} to explore samples from the joint distribution ordered by the wait time at that shop.} 
\end{figure}

In order to investigate whether interactivity can improve rationality or comprehension, we propose interactive variants of boxplots and HOPs that allow the user to focus on self-defined subspaces of the joint distribution. Interactive Boxplots allow users to explore a model by conditioning the joint distribution on one or more parameters. A mouse-click conditions the model on a small range of values of the selected parameter around the click location and updates the boxplots of all other parameters according to their marginals under the conditional distribution\footnote{Video hosted on Google Drive for double-blind review \url{https://drive.google.com/file/d/1VXoLeeqcGuDHMKbh5bNjGT05q25o6pVm/view}}. An example conditioning on one variable is shown in Fig.~\ref{fig:interactive_vis} (left). Additionally, the width of the selected parameter is proportional to the probability of the condition under the joint. Thereby the user can explore potential pair-wise correlations and complex conditionals. We propose Ballistic Hypothetical Outcome Plots (BHOPs), which allow the user to guide exploration of the joint distribution by clicking on one parameter of interest to trigger a “pseudo-physical” sequence of draws\footnote{Video hosted on Google Drive for double-blind review \url{https://drive.google.com/file/d/1HAZBo8bY2oOwHtXMuvlqWQ_Fww2LwnSI/view}}. The user can explore potential pairwise correlations and condition the joint distribution on a subset of values for one variable by “pinging” variables of interest (see Fig.~\ref{fig:interactive_vis} (right)).

\subsection{Tasks}

We were interested in exploring the suitability of the proposed visualisations as a generic interface to Bayesian models, and thus evaluate their effect on user responses across a large portion of the Query Space (see Section~\ref{sec:query_space}). A set of 24 queries were generated covering the outer product of two levels of Observability (observable, latent), three Quantities (value, confidence, and id), two levels of Conditioning (posterior, and posterior with additional side-information), and two variations of each query type. One variation of a query with additional side-information concerning the confidence in the latent ID of a variable provided the \textbf{Context} \emph{A friend tells you she just waited for her sandwich exactly 6 minutes, but won't tell you which sandwich shop she visited. A mobile app shows you a long-term distribution of wait times across nearby sandwich shops as in the plot. You also have the additional information (not represented in the plot) that it currently takes between 3.5 and 4.0 minutes to get served at Sandys.} and posed the \textbf{Query} \emph{Considering this additional information, how confident are you that your friend got her sandwich from Mesys?}. The additional information from a trustworthy source should be used to further condition the model on queuing delays between 3.5 and 4.0 minutes at Sandys. While this is not possible with non-interactive boxplots, it affords visual filtering of samples shown with HOPs, and interacting with the visual representation of the distribution of queuing delays at Sandys in Interactive Boxplots and Ballistic Hypothetical Outcome Plots. The full list of questions can be seen in the Supplementary Material\footnote{user\_tasks.txt}.

\subsection{Design}

A factorial design was chosen to measure effects of the four Model Visualisations on all 24 tasks with 96 conditions in total. A pilot study established that only 48 conditions could feasibly be evaluated with each participant within one hour of participant time including introduction and debriefing. As a result, each variant of each query type was only paired with a random subset of two visualisation types per participant. 

The interactive variants of boxplots and HOPs intentionally introduce only minimal changes to the visual appearance of their non-interactive counterparts. As a consequence, it is not immediately obvious to the user whether the specific visualisation they are presented with in each task of a randomized sequence affords interaction. We therefore separated all tasks in which interactive and non-interactive visualisations were presented, randomized the order of these groups, and the order of tasks within each group. Each group was further preceded by a message highlighting whether the subsequent set of tasks afforded interaction.

\subsection{Procedure}

The user study was conducted in the presence of a researcher on a laptop s/he provided. Participants were provided with an information sheet upon arrival and asked to sign a consent form. They received a £10 Amazon voucher for taking part and were promised to receive an additional £50 Amazon voucher if they obtained the highest cumulative reward among all participants. They were seated in front of a table. An introduction to the study, explaining the types of Queries, Model Visualisations and Answer Inputs, and describing the Task Feedback was delivered through the client interface and gave participants the opportunity to interact with each of these UI elements prior to commencing the study. Participants were offered clarifications at the end of the introduction, were reassured that they are responsible for advancing through the study and that they could take breaks as needed. They received task feedback after each task, and their cumulative reward was shown in the top right corner of the display at all times. They were informed when they advanced from the set of interactive visualisations to non-interactive ones or vice versa. After all tasks were completed, they were invited to provide qualitative feedback and given the opportunity to ask questions in a short debriefing discussion.

\subsection{Participants}

Participants were recruited via mailing lists and Yammer from former and present student populations, administrative, and academic staff across the University. The study was approved in advance by the University Ethics Committee. 22 volunteers took part in the user study, providing a total of $22*48=1056$ user responses.

\section{Analysis Workflow}
\label{sec:analysis}

This Section illustrates the proposed analysis workflow on the data collected from the case study in Section \ref{sec:case_study}. 

\subsection{User Study Results}

\subsubsection{Was the task difficulty well calibrated?} 
\begin{figure}
    \centering
    \includegraphics[width=\textwidth]{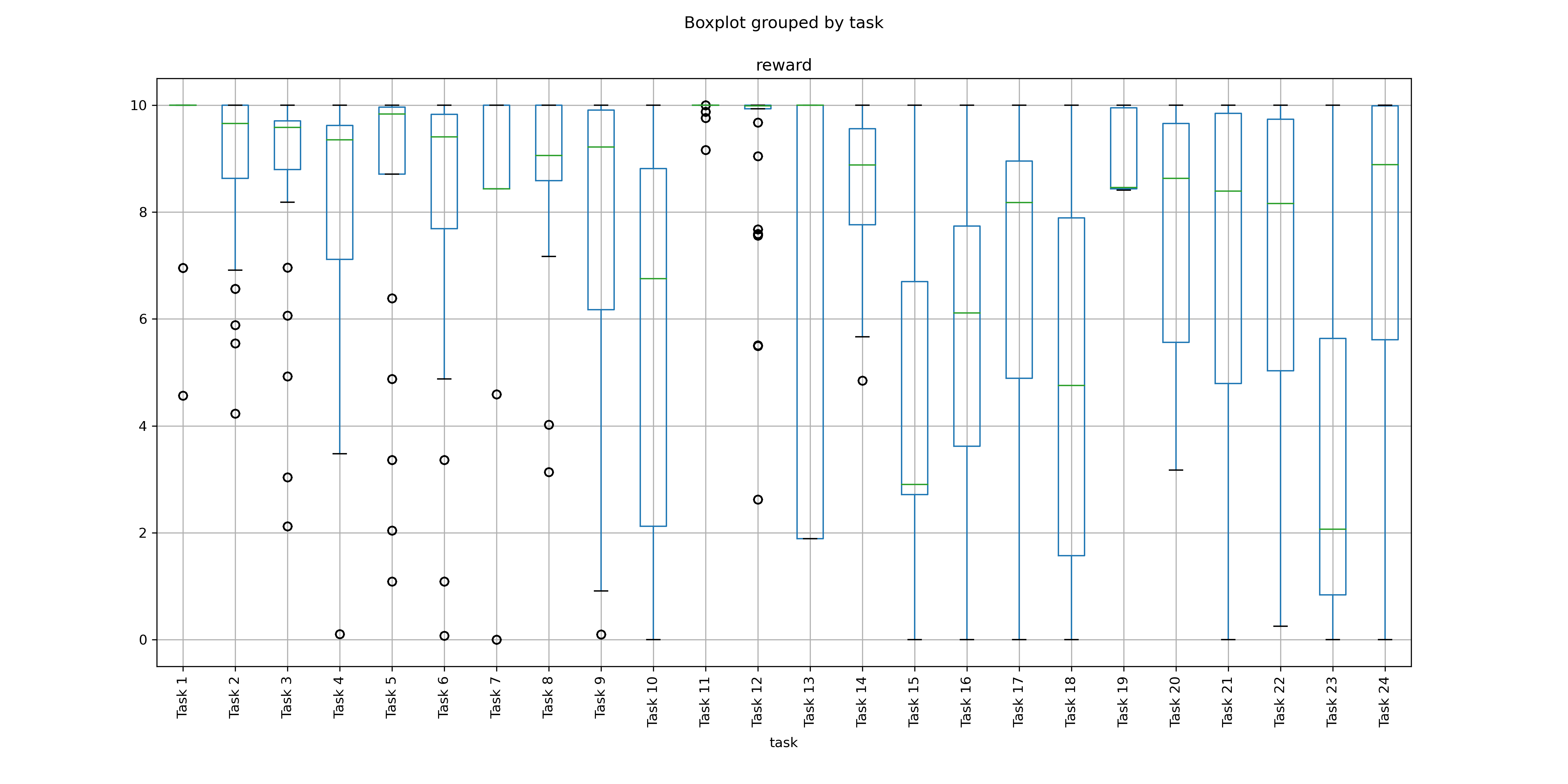}
    \caption{Stratified sample statistics of rewards $\in [0, 10]$ by study task, where stratification was performed across Model Visualisations. Most tasks appear sufficiently challenging as sample rewards deviate considerably from the optimal reward. Tasks 1, 11, and 12 appear too easy after post-hoc analysis.}
    \label{fig:task_rewards}
    \Description{Stratified sample statistics of rewards $\in [0, 10]$ by study task, where stratification was performed across Model Visualisations. Most tasks appear sufficiently challenging as sample rewards deviate considerably from the optimal reward. Tasks 1, 11, and 12 appear too easy after post-hoc analysis.} 
\end{figure}

The difficulty of study tasks should ideally be calibrated such that user responses deviate sufficiently from the optimal task response, confirming that a task was not too easy, and such that user responses are significantly different from a random response, confirming that a task was not too difficult. Note that only limited calibration can be performed a priori as it requires user responses. The sample of rewards obtained by participants for each study task, stratified across Model Visualisations, is illustrated in Fig.~\ref{fig:task_rewards}. Almost all tasks are sufficiently challenging as sample rewards deviate considerably from the optimal reward. Tasks 1, 11, and 12 appear potentially too easy in hindsight, as the vast majority of participant responses on these tasks were close to optimal. A random agent's behavior was simulated computationally by sampling 1000 responses uniformly at random from the space of permissible responses for each task. The corresponding rewards were compared to participants' rewards. A Mann-Whitney U test with Bonferroni correction for multiple comparisons rejected the null-hypothesis ($p_i \le \frac{\alpha=0.05}{m=24}$) simultaneously for all tasks. This result suggests that users in aggregate responded to no task at random. Note, however, that this comparison provides a weak lower bound on task difficulty as a random agent completely disregards the query and the visualised information of the model.

\subsubsection{Does visual animation have an effect on rationality or comprehension?} We investigate the potential effect of visual animation on user responses as measured by task reward. We perform a within-subject analysis of differences in rewards obtained on the same task by each user under an animated and a non-animated condition. Specifically, we consider (B)HOPs as animated and (interactive) boxplots as not animated. Paired samples were stratified by task and aggregated separately across \emph{rationality} questions and \emph{comprehension} questions. The 50\% central interval (CI) of differences in rewards observed in response to rationality questions $[-0.38, 0.84]$ and to comprehension questions $[-1.36, 0.85]$ overlaps with the null hypothesis, thus animation did not have an effect on rewards in this study. There was also no effect on the differences in time until an answer was acknowledged, with 50\%-CI of $[-11.50, 11.33]$ and $[-16.84, 10.95]$ for rationality and comprehension questions, respectively.

\subsubsection{Does interactivity have an effect on rationality or comprehension?} Analogously, we investigate the effect of interactivity by analysing paired samples of responses under an interactive and non-interactive condition. We consider interactive boxplots and BHOPs as interactive, and boxplots and HOPs as non-interactive. The central 50\%-CI of differences in rewards observed in response to rationality questions $[-0.92, 1.33]$ and to comprehension questions $[-0.69, 0.85]$ also overlap with the null hypothesis, indicating that interactivity also had no significant effect on rewards in this study. Similarly, there was no effect of interactivity on response time with 50\%-CI of $[-15.83, 11.90]$ and $[-23.55, 14.63]$ for rationality and comprehension questions, respectively.

\subsection{Discussion}

Jointly, these results highlight the need for quantitative evaluation to test whether knowledge gained in other scenarios where interactive and animated visualisations have been found to be useful readily applies to the interpretation of Bayesian models. By no means do the results reported here suggest otherwise. Instead, one might speculate that previously identified challenges with communicating uncertainty are exacerbated by the potentially high dimensionality and complex covariance structure represented by Bayesian models, that effects are limited to a smaller query subspace, or that effect sizes of subtle changes to existing visualisations of uncertainty are too small to be measured with lab-based user studies as suggested by the results of \citet{micallef2012}. The absence of a measured effect may also be related to other confounding factors \cite{hullman_error} such as the users' context, the framing of the task query, numerical information present in the task context, a mismatch between users' reported subjective confidence and elicited statistical confidence, or participants' employed heuristics to simplify judgements under uncertainty. More radically innovative interactive tools that communicate aspects of a Bayesian model's structure might also be more supportive of rational decision making and model comprehension \cite{taka2020}. Each of these hypotheses can be tested with the proposed evaluation protocol and the software framework presented in this paper by making systematic changes to the study design. In the context of assessing task difficulty we introduced computational modelling, with a very naive behavior model, as a tool for interpreting user responses. In future work we will investigate this approach further to explore, for example, how user responses compare to an agent whose decisions are based on restricted views of the joint distribution such as the global mean, disregarding all information on uncertainty, or unconditional marginal distributions.

\section {Conclusion}

While reasoning about uncertainty is challenging for people, and communicating uncertainty is difficult, systematic research in this area is becoming increasingly important as day-to-day decisions and scientific discoveries are driven by probabilistic models. Standardised evaluation protocols support meta-studies, reproducibility and measuring progress of the research community as a whole, but are currently lacking in the field of probabilistic model visualisation. With the proposed protocol and software framework, we hope to contribute to this community effort.

The evaluation protocol delineates the query space for Bayesian models and proposes one semantic mapping as a tool to define user study tasks. It draws the distinction between objectives for evaluating decision making and model comprehension, and formalizes each for quantitative evaluation. We identified a gap in available user input controls for communicating uncertainty in categorical decisions and designed the MultiBet in response. To further advance the standardisation effort, we developed a customisable software framework for user studies on probabilistic model visualisations that can be extended easily to include further input controls \cite{greis2017b} and visualisation types \cite{taka2020}. We reported on a user study to illustrate the process of conducting a quantitative user study with this framework, and to test the research hypotheses whether animation or interaction has an effect on rationality and comprehension, which were motivated by findings in the literature. The quantitative evaluation results did not support either of these hypotheses, highlighting the need for scrutinizing design hypotheses with quantitative user studies.

\begin{acks}
This work was supported by the \grantsponsor{EPSRC} EPSRC Project: EP/R018634/1 \grantnum{EP/R018634/1}{Closed-Loop Data Science for Complex, Computationally- and Data-Intensive Analytics}.
\end{acks}

\bibliographystyle{ACM-Reference-Format}
\bibliography{references}

\section*{Appendix}

We also performed a between-subject analysis of responses, exploring in more detail differences in accuracy on task level. Using bootstrap with 1000 samples, we estimated the mean and standard deviation of accuracy per task and per condition, and estimated the effect size as the difference in sample means scaled by the square root of the average standard deviation within each pair of conditions.

\begin{figure}
    \centering
    \includegraphics[width=\textwidth]{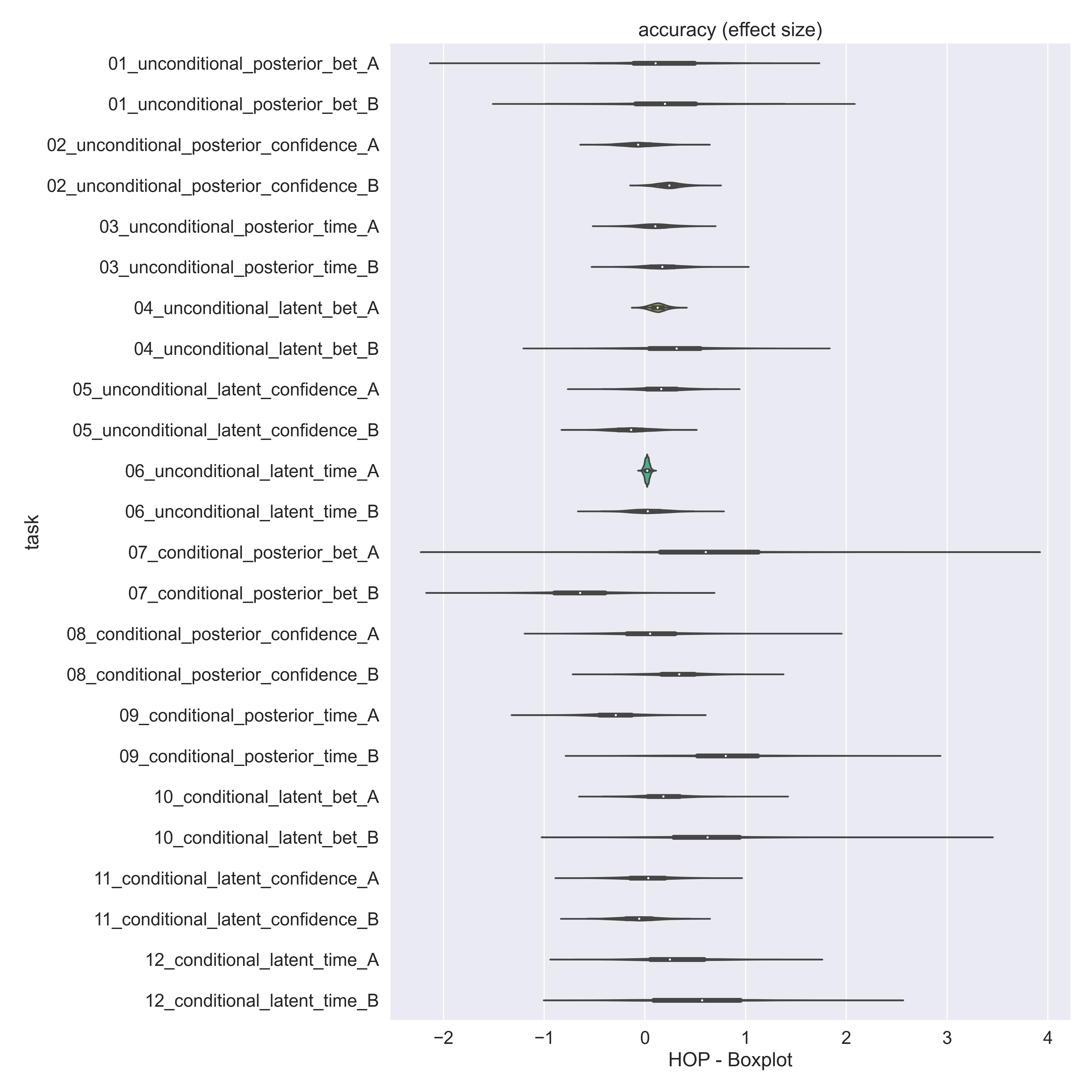}
    \caption{Effect size of differences in accuracy between HOP and Boxplot conditions.}
\end{figure}

\begin{figure}
    \centering
    \includegraphics[width=\textwidth]{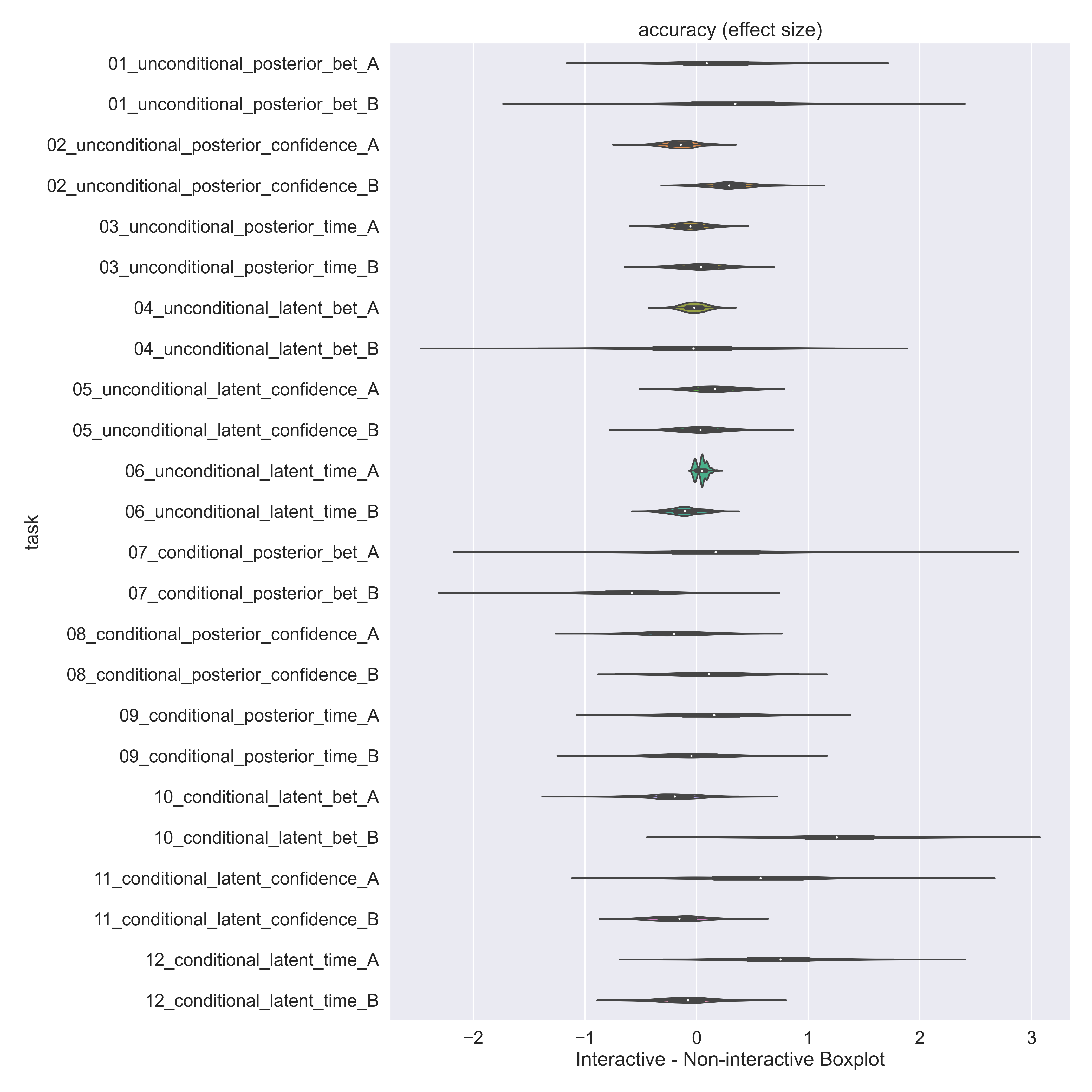}
    \caption{Effect size of differences in accuracy between interactive and non-interactive Boxplot conditions.}
\end{figure}

\begin{figure}
    \centering
    \includegraphics[width=\textwidth]{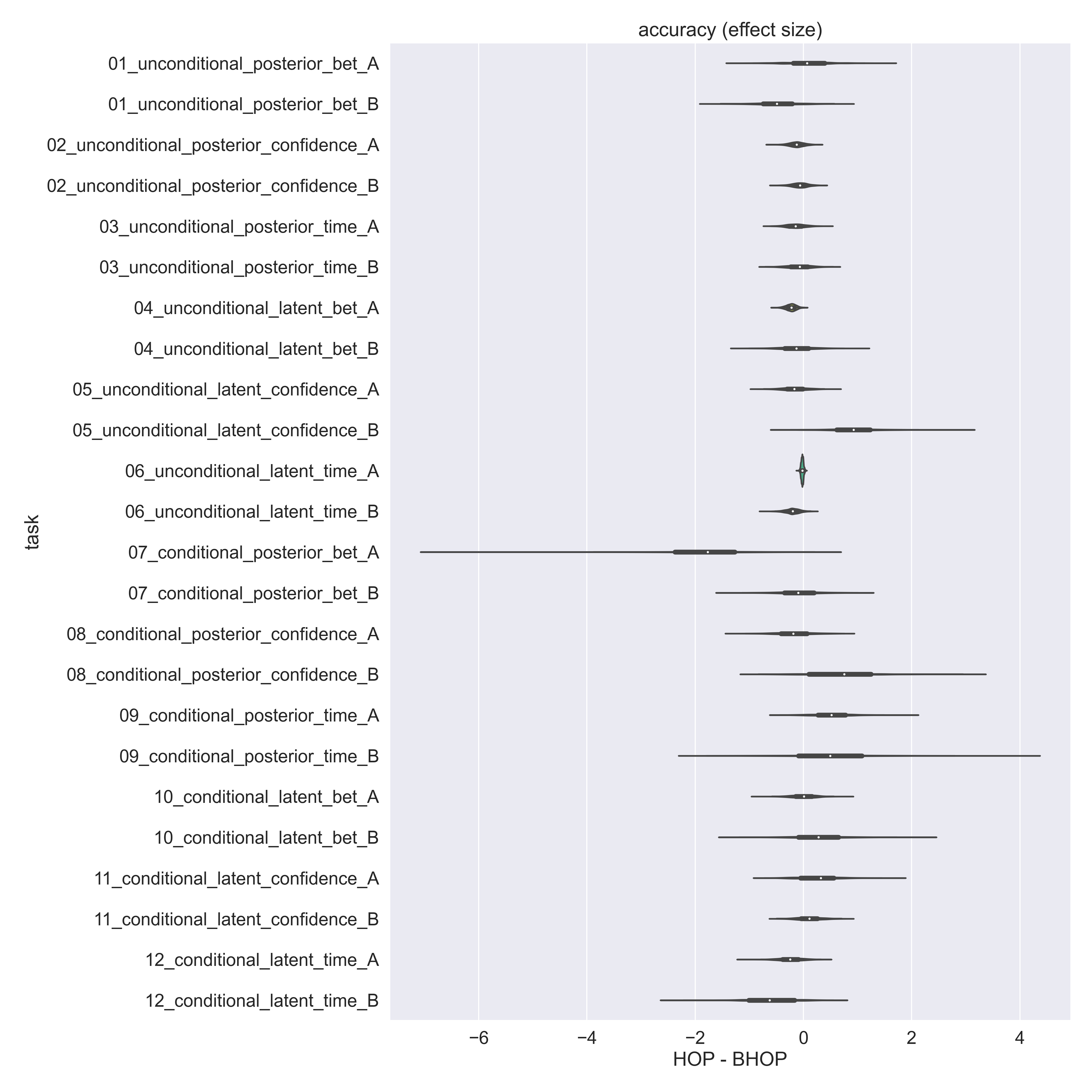}
    \caption{Effect size of differences in accuracy between BHOP and HOP conditions.}
\end{figure}

\end{document}